# Space-Time Codes from Structured Lattices

K. Raj Kumar and Giuseppe Caire

### Abstract

We present constructions of Space-Time (ST) codes based on lattice coset coding. First, we focus on ST code constructions for the short block-length case, i.e., when the block-length is equal to or slightly larger than the number of transmit antennas. We present constructions based on dense lattice packings and nested lattice (Voronoi) shaping. Our codes achieve the optimal diversity-multiplexing tradeoff of quasi-static MIMO fading channels for any fading statistics, and perform very well also at practical, moderate values of signal to noise ratios (SNR). Then, we extend the construction to the case of large block lengths, by using trellis coset coding. We provide constructions of trellis coded modulation (TCM) schemes that are endowed with good packing and shaping properties. Both short-block and trellis constructions allow for a reduced complexity decoding algorithm based on minimum mean squared error generalized decision feedback equalizer (MMSE-GDFE) lattice decoding and a combination of this with a Viterbi TCM decoder for the TCM case. Beyond the interesting algebraic structure, we exhibit codes whose performance is among the state-of-the art considering codes with similar encoding/decoding complexity.

The authors are with the Department of Electrical Engineering - Systems, University of Southern California, Los Angeles, CA 90089, USA ({`rkkrishn,caire`}@usc.edu).

The material in this paper was presented in part at the IEEE International Symposia on Information Theory ISIT-2006 in Seattle, USA and ISIT-2007 in Nice, France.

This work has been partially funded by a gift from ST Microelectronics, NSF Grant No. CCF-0635326, the 2006 Okawa Foundation Research Grant and by an Oakley fellowship from the Graduate School at USC.

# I. INTRODUCTION

The quasi-static, frequency-flat fading (complex) multiple-input multiple-output (MIMO) channel with $M$ transmit and $N$ receive antennas and coding block-length $T$ channel uses is described by

$$\mathbf{Y}^c = \mathbf{H}^c \mathbf{X}^c + \mathbf{W}^c, \tag{1}$$

where $\mathbf{X}^c$ denotes the $M \times T$ transmitted codeword matrix drawn from a space-time (ST) code $\mathfrak{X}$, $\mathbf{Y}^c$ is the $N \times T$ received signal matrix, $\mathbf{H}^c$ is the $N \times M$ channel matrix and $\mathbf{W}^c$ is the $N \times T$ noise matrix. The entries of the channel matrix $\mathbf{H}^c$ are assumed to be constant over a block length of $T$ channel uses and the entries of $\mathbf{W}^c$ are independent and identically distributed complex Gaussian with zero mean and unit variance, i.e., i.i.d. $\mathcal{CN}(0, 1)$. The results of this paper will hold for arbitrary channel fading statistics, but we will use the standard i.i.d. Rayleigh fading model for our simulations, in which case the entries of $\mathbf{H}^c$ are i.i.d. $\mathcal{CN}(0, 1)$. The input constraint

$$\mathbb{E}\|\mathbf{X}^c\|_F^2 \leq T \text{ SNR} \tag{2}$$

is enforced, where $\mathbb{E}(\cdot)$ denotes the expectation operator and SNR takes on the meaning of the *transmit* signal-to-noise ratio (total transmit energy per channel use over the noise power spectral density). The channel matrix $\mathbf{H}^c$ is assumed to be known perfectly at the receiver but not at the transmitter.

The use of ST codes over MIMO channels is known to provide two kinds of benefits: better reliability through diversity gain, and higher data rates in terms of multiplexing gain. The diversity-multiplexing tradeoff (DMT) (see [9] for the definition and details) captures in a succinct and elegant way the tradeoff between these two quantities in the high signal to noise ratio (SNR) regime. The DMT specifies the maximum possible diversity that can be obtained at each possible value of multiplexing gain, and has become a standard performance metric to evaluate ST schemes, and a tool to compare different ST schemes.

Families of codes that achieve the DMT of MIMO fading channels have been proposed. Perhaps the most notable in terms of performance and generality are Lattice ST (LaST) codes and codes obtained from cyclic division algebras (CDA).

An ensemble of randomly generated LaST codes was shown to be DMT optimal under minimum mean squared error generalized decision feedback equalizer (MMSE-GDFE) lattice decoding for $T \geq M+N-1$ [1]. In this case, DMT optimality is shown in a random coding sense (i.e., with respect to error probability averaged over the random lattice ensemble) and for the Rayleigh i.i.d. fading statistics.

Families of carefully constructed CDA codes enjoy the so-called *non-vanishing determinant* (NVD) property (to be defined subsequently), which in turns implies that these codes, under ML decoding, achieve the optimal DMT in a universal sense, i.e., over any channel fading statistics [2]. Codes achieving the optimal DMT over any fading statistics are called "approximately universal" in [3]. Furthermore, these codes allow for minimum block length, i.e., there exist optimal codes for all $T \geq M$ [2].

In some sense, the present work may be thought of as a confluence of these two approaches. We construct codes that retain desirable properties from both families: not only are they are non-random explicit constructions from CDAs, but they also employ the nested lattice construction that enables shaping gains and the reduced complexity MMSE-GDFE lattice decoding akin to the LaST codes.

The DMT captures the optimal performance for high SNR. Following [1], [2], attention has shifted towards constructing ST codes that not only achieve the DMT, but also perform well at finite (practical) values of SNR. For example, generating codes at random from the ensemble of [1] yields typically performances that stay at 1 to 3 dB from outage probability (that can be regarded an effective "quasi-lower bound" on the performance of any code at meaningful SNR, i.e., for probability of block error not too large (say, $\leq 10^{-1}$)). In this perspective, the first part our this work presents a construction of *structured* LaST (S-LaST) codes[1] that achieve the DMT and perform well at finite SNR, for small to moderate block-lengths (i.e., $T$ is equal to or slightly larger than $M$). In the second part of the paper we turn to the case of large block lengths $T \gg M$. This is motivated by the fact that in practical wireless communication systems, information is encoded and sent over the channel in packets, together with training symbols, protocol information, and guard intervals. Therefore, packets cannot be too small, for otherwise the overhead would be a large part of the overall capacity. We target the case where data packets span a number of channel uses $T$ considerably larger than the number of transmit antennas $M$, but nevertheless smaller than a fading coherence interval. Then, the fading channel is constant over the whole codeword of duration $T$ channel uses.

Unfortunately, the LaST and/or CDA constructions do not generalize, in practice, to $T \gg M$ since the decoding complexity grows rapidly with $T$. Furthermore, with constructions such as those in [1], [2] it is not clear how to exploit the large block length to obtain codes with improved coding gain. Therefore, the challenge here is to design ST codes for large $T$ that have good coding gain and low decoding complexity. In this regard, the authors in [21] have proposed a trellis coded modulation (TCM) scheme based on partitions of the Golden code [11]. For prior work on ST TCM, see [18], [19]. Building on these ideas, we propose a general technique for the construction of ST-TCM schemes with good coding and shaping gains. These codes can be decoded using the Viterbi Algorithm where the branch metrics are computed using a low complexity MMSE-GDFE lattice decoder. We show construction examples based on the Gosset lattice $E_8$ and lattices drawn from the Golden+ algebra [12] that yield, to the best of the authors' knowledge, the current state-of-the art performance among codes with similar encoding/decoding complexity.

[1]We use the term "structured" to distinguish these codes from the random lattice approach of [1].

In Section II we review LaST codes and ST codes from CDAs, as these form the two main ingredients for our construction. We also review some concepts relating to lattice packings that will be used subsequently. Code design for the short block-length case is presented in Section III, and Section IV deals with the construction of TCM schemes. Simulations results are provided alongside each construction, and illustrate the effectiveness of the constructions.

## II. BACKGROUND

### A. Lattice Space-Time (LaST) codes

An $n$-dimensional real lattice $\Lambda$ is a discrete additive subgroup of $\mathbb{R}^n$ defined as $\Lambda = \{\mathbf{G}\mathbf{u} \ : \ \mathbf{u} \in \mathbb{Z}^n\}$, where $\mathbf{G}$ is the $n \times n$ (full-rank) real generator matrix of $\Lambda$. The fundamental Voronoi cell of $\Lambda$, denoted as $\mathcal{V}(\Lambda)$, is the set of points $\mathbf{x} \in \mathbb{R}^n$ closer to zero than to any other point $\boldsymbol{\lambda} \in \Lambda$. The fundamental volume of $\Lambda$ is

$$V_f(\Lambda) \triangleq V(\mathcal{V}(\Lambda)) = \int_{\mathcal{V}(\Lambda)} d\mathbf{x} = \sqrt{\det(\mathbf{G}^T\mathbf{G})}.$$

An $n$-dimensional lattice code $\mathcal{C}(\Lambda, \mathbf{u}_0, \mathcal{R})$ is the finite subset of the lattice translate $\Lambda + \mathbf{u}_0$ inside the shaping region $\mathcal{R}$, i.e., $\mathcal{C} = \{\Lambda + \mathbf{u}_0\} \cap \mathcal{R}$, where $\mathcal{R}$ is a bounded measurable region of $\mathbb{R}^n$.

LaST codes are more easily illustrated by considering the real vectorized channel model equivalent to (1),

$$\mathbf{y} = \mathbf{H}\mathbf{x} + \mathbf{w}, \tag{3}$$

where $\mathbf{x} \in \mathbb{R}^{2MT}$ and $\mathbf{y}, \mathbf{w} \in \mathbb{R}^{2NT}$ denote respectively the vector equivalents of $\mathbf{X}^c, \mathbf{Y}^c$ and $\mathbf{W}^c$ obtained by separating real and imaginary part and by stacking columns, and where $\mathbf{H} = \mathbf{I}_T \otimes \begin{bmatrix} \mathrm{Re}(\mathbf{H}^c) & -\mathrm{Im}(\mathbf{H}^c) \\ \mathrm{Im}(\mathbf{H}^c) & \mathrm{Re}(\mathbf{H}^c) \end{bmatrix}$, according to the well-known construction as in [1]. We say that an $M \times T$ space-time coding scheme $\mathcal{X}$ is a full-dimensional LaST code if it's vectorized (real) codebook (corresponding to the channel model in (3)) is a lattice code $\mathcal{C}(\Lambda, \mathbf{u}_0, \mathcal{R})$, for some $n$-dimensional lattice $\Lambda$, translation vector $\mathbf{u}_0$, and shaping region $\mathcal{R}$, where $n = 2MT$. Given the equivalence of the real vector and the complex matrix representation of $\mathcal{X}$, we shall not distinguish between them explicitly and write simply $\mathcal{X} = \mathcal{C}(\Lambda, \mathbf{u}_0, \mathcal{R})$. Any linear-dispersion ST code, including the constructions of [2], can be represented as a LaST code, for a suitable shaping region. For later use, we define the lattice quantization function as

$$Q_\Lambda(\mathbf{y}) \triangleq \arg\min_{\boldsymbol{\lambda} \in \Lambda} |\mathbf{y} - \boldsymbol{\lambda}|$$

and the modulo-lattice function

$$[\mathbf{y}] \bmod \Lambda = \mathbf{y} - Q_\Lambda(\mathbf{y}).$$

We also define the notion of a non-vanishing determinant (NVD) for an infinite LaST code (i.e., disregarding the shaping region $\mathcal{R}$) as follows. A LaST code has the NVD property if and only if the minimum determinant corresponding to its infinite lattice $\Lambda$ is bounded away from zero by a constant independent of SNR, i.e.,[2]

$$\min_{\substack{\Delta \mathbf{X}^c = \mathbf{X}_i^c - \mathbf{X}_j^c, \\ \mathbf{x}_i \neq \mathbf{x}_j, \ \mathbf{x}_i, \mathbf{x}_j \in \Lambda + \mathbf{u}_0}} \det\left[\Delta \mathbf{X}^c (\Delta \mathbf{X}^c)^{\mathsf{H}}\right] \ \dot{\geq} \ \mathrm{SNR}^0.$$

Notice that since $\Lambda$ is a lattice, this is equivalent to

$$\min_{\mathbf{x} \in \Lambda + \mathbf{u}_0} \det\left[\mathbf{X}^c (\mathbf{X}^c)^{\mathsf{H}}\right] \ \dot{\geq} \ \mathrm{SNR}^0.$$

### B. ST Codes from CDA

For a detailed exposition of ST codes from CDA, we refer the reader to [24], [2] and references therein. We provide a very brief review in the sequel. Let $\mathbb{Q}$ denote the field of rational numbers and $\imath \triangleq \sqrt{-1}$. Set $\mathbb{F} = \mathbb{Q}(\imath)$. The construction of a CDA calls for the construction of an $n$-degree cyclic Galois extension $\mathbb{L}/\mathbb{F}$ with generator $\sigma$. Then a CDA $D(\mathbb{L}/\mathbb{F}, \sigma, \gamma)$ with center $\mathbb{F}$, maximal subfield $\mathbb{L}$ and index $n$ is the set of all elements of the form $\sum_{i=0}^{n-1} z^i \ell_i$, where $z$ is an indeterminate satisfying $\ell z = z\sigma(\ell) \ \forall \ \ell \in \mathbb{L}$ and $z^n = \gamma$. The element $\gamma$ needs to be a properly chosen *non-norm element* in order to ensure that $D$ is a division algebra, see [24], [2] for details. Every element in the CDA can be associated with an $n \times n$ matrix through the *left regular representation*, which is of the form

$$\begin{bmatrix} \ell_0 & \gamma\sigma(\ell_{n-1}) & \gamma\sigma^2(\ell_{n-2}) & \ldots & \gamma\sigma^{n-1}(\ell_1) \\ \ell_1 & \sigma(\ell_0) & \gamma\sigma^2(\ell_{n-1}) & \ldots & \gamma\sigma^{n-1}(\ell_2) \\ \vdots & \vdots & \vdots & \ddots & \vdots \\ \ell_{n-1} & \sigma(\ell_{n-2}) & \sigma^2(\ell_{n-3}) & \ldots & \sigma^{n-1}(\ell_0) \end{bmatrix}, \tag{4}$$

---

[2]We make use of the exponential equality notation from [9], defined as

$$a \doteq \rho^{-b} \Leftrightarrow b = -\lim_{\rho \to \infty} \frac{\log a}{\log \rho}.$$

The notations $\dot{\geq}$ and $\dot{\leq}$ are defined similarly.

where $\ell_i \in \mathbb{L}$. The trace and determinant of the above matrix are respectively defined to be the *reduced trace* $\mathrm{tr}_r(\cdot)$ and *reduced norm* $N_r(\cdot)$ of the element it represents. The ST code with $M = T = n$ is a finite collection of matrices of the above form, scaled to satisfy the power constraint in (2). Choosing $\gamma \in \mathbb{Z}[\imath]$ and restricting the $\ell_i$ to belong to the ring of integers $\mathcal{O}_\mathbb{L}$ of $\mathbb{L}$ bestows the NVD property on the ST code. One such choice for the $\ell_i$ corresponds to choosing

$$\ell_i = \sum_{k=1}^{n} e_{i,k}\beta_k, \quad e_{i,k} \in \mathcal{A}_{\mathrm{QAM}}, \tag{5}$$

with $\mathcal{A}_{\mathrm{QAM}} = \{a + \imath b \ | \ -Q + 1 \le a, b \le Q - 1, \ a, b \ \text{ odd } \}$, and where $\beta_k, k = 1, 2, \ldots, n$ is an integral basis (i.e., a basis as a module) for $\mathcal{O}_\mathbb{L}/\mathcal{O}_\mathbb{F}$. More generally, we could choose $\{\beta_k\}_{k=1}^{n}$ to constitute an $\mathcal{O}_\mathbb{F}$-basis for any ideal $\mathcal{I} \subseteq \mathcal{O}_\mathbb{L}$. In this case, $|\mathfrak{X}| = Q^{2n^2}$. The results of [2], [3] show that codes derived from CDA with NVD are *approximately universal*.

In the recent work [12], ST codes are obtained from *maximal orders* in CDAs. For the sake of later use, a brief review follows. A $\mathbb{Z}[\imath]-order$ in an $\mathbb{F}-$algebra $D$ is a subring $O$ of $D$, having the same identity element as $D$, and such that $O$ is a finitely generated module over $\mathbb{Z}[\imath]$ and generates $D$ as a linear space over $\mathbb{F}$.

An order $O$ is called *maximal* if it is not properly contained in any other $\mathbb{Z}[\imath]-$order. The discriminant of a $\mathbb{Z}[\imath]-$order $O$ is computed as $d(O/R) = \det([\mathrm{tr}_r(b_i b_j)]_{i,j=1}^{m})$, where $\{b_1, \ldots, b_m\}$ is any $\mathbb{Z}[\imath]-$basis of $O$.

All maximal orders of a CDA share the same value of the discriminant, and also have the smallest possible discriminant among all orders within a given CDA. An important property of elements of an order of a CDA $D(\mathbb{L}/\mathbb{F}, \sigma, \gamma)$ is that their reduced norm (i.e., the determinant of their matrix representation) is an element of the ring of integers $\mathcal{O}_\mathbb{F} = \mathbb{Z}[\imath]$ of the center $\mathbb{F}$. This property ensures that ST codes carved out of orders in suitably constructed CDAs are endowed with the NVD property. The choice of a subset of elements of $D$ corresponding to (5) amounts to choosing a particular order $O$ known as the *natural order*.

It is established in [12] that the discriminant of an order in a CDA is directly proportional to the fundamental volume of the ensuing lattice (they are in fact equal for the case when the center of the CDA is $\mathbb{F} = \mathbb{Q}(\imath)$). Therefore, in order to maximize the energy efficiency of the code, a sensible design guideline is to use the maximal order of the CDA to derive ST codes, owing to them having the minimum possible discriminant. All previous constructions of ST codes from CDAs, including the ones in [24], [2], [4], [11], [5] have used the natural order, which is not guaranteed to be maximal in general.

As an illustration of the technique, the authors in [12] construct a $2 \times 2$ ST code derived from the maximal order of a CDA named the Golden+ Algebra ($\mathcal{GA}+$), whose minimum determinant improves upon that of previously known constructions. We will revisit this construction subsequently in Section. III, and use it to construct some of our examples.

### C. Lattice Packings

The classical sphere packing problem is to find how densely a large number of identical spheres can be packed together in $n$-dimensional space. A packing is called a lattice packing if it has the property that the set of centres of the spheres forms a lattice in $n$-dimensional space. An excellent reference for this area is the book by Conway and Sloane [6].

The *density* $\Delta$ of a lattice packing is given by

$$
\begin{aligned}
\Delta \quad &\triangleq \quad \text{Proportion of space that is occupied by the spheres} \\
&= \quad \frac{\text{volume of one sphere}}{V_f(\Lambda)}.
\end{aligned}
$$

A related quantity is the *center density* $\delta$, given by

$$\delta = \frac{\Delta}{V_n},$$

where $V_n$ is the volume of an $n$-dimensional sphere of radius 1, given by

$$V_n = \frac{\pi^{n/2}}{(n/2)!} = \frac{2^n \pi^{(n-1)/2}((n-1)/2)!}{n!}$$

(the second form avoids the use of $(n/2)!$ when $n$ is odd). A related parameter is the *fundamental coding gain* $\gamma_c(\Lambda)$, defined as:

$$\gamma_c(\Lambda) \triangleq 4\delta^{2/n} = \frac{d_{\min}^2(\Lambda)}{V(\Lambda)^{2/n}}, \tag{6}$$

where $d_{\min}(\Lambda)$ denotes the minimum distance of the lattice $\Lambda$. It is evident from the definition that the fundamental coding gain is a normalized measure of the density of the lattice. Further, the fundamental coding gain also possesses the desirable properties of being dimensionless, and invariant to scaling and any orthogonal transformation (rotation) [8]. For the cubic lattice, $\gamma_c(\mathbb{Z}^n) = 1$.

The problem of finding dense packings (i.e., those with high values of $\gamma_c(\Lambda)$) in $n$-dimensional space has a long and interesting history. In two dimensions, Gauss proved that the hexagonal lattice is the densest plane lattice packing, and in 1940, L. Fejes Tóth proved that the hexagonal lattice is indeed the densest of all possible plane packings. In 1611, the German astronomer Johannes Kepler stated that no packing in three dimensions can be denser than that of the face-centered cubic (f.c.c.) lattice arrangement which fills about 0.7405 of the available space. It took mathematicians some 400 years to prove him right, with Thomas Hales proving the conjecture in 1998 (Gauss showed in 1821 that the f.c.c. lattice is the densest possible *lattice* packing in three dimensions). The densest possible lattice packings are known for all dimensions $n \leq 8$. The checkerboard lattices $D_4$ and $D_5$ are the densest possible lattice packings in 4 and 5-dimensions respectively while Gosset's root lattices $E_6, E_7$ and $E_8$ are optimal among lattice packings in $6, 7$ and 8-dimensions. It is also known that the densest lattice packings in dimensions 1 to 8 are unique. Although not proven, it seems likely that Coxeter-Todd lattice $K_{12}$, the Barnes-Wall lattice $\Lambda_{16} \cong BW_{16}$ and the Leech lattice $\Lambda_{24}$ are the densest lattices in dimensions $12, 16$ and 24 respectively [6]. Tables of the best known lattice packings in $n$-dimensions are available in the literature [6] and in the online catalogue of lattices [7].

For later use, we define a lattice $\Lambda$ with generator matrix $\mathbf{G}$ to be an *integral lattice* if the Gram matrix $\mathbf{A} \triangleq \mathbf{G}^\mathsf{T}\mathbf{G}$ has integer entries. It turns out that many of the best known lattices in terms of packing belong to this class, when suitably scaled.

## III. The Structured LaST Code Construction

This section deals with code design for the case of short block-lengths, i.e., $T$ is equal to or slightly larger than $M$. Before we present the construction, we first explore the LaST formulation of space-time codes derived from CDA.

### A. CDA ST Codes as Lattice Codes

We will illustrate the equivalent lattice structure with an example of a $2 \times 2$ ST code derived from CDA. From (4), any codeword matrix is of the form

$$\mathbf{X}^c = \begin{bmatrix} \ell_0 & \gamma\sigma(\ell_1) \\ \ell_1 & \sigma(\ell_0) \end{bmatrix}.$$

The real vector corresponding to $\mathbf{X}^c$ in the equivalent channel model of (3) is given by

$$\mathbf{x} = \left[ \mathrm{Re}(\mathbf{x}^c)^\mathsf{T} \mathrm{Im}(\mathbf{x}^c)^\mathsf{T} \right]^\mathsf{T},$$

where

$$\mathbf{x}^c = [\ell_0 \ \ell_1 \ \gamma\sigma(\ell_1) \ \sigma(\ell_0)]^\mathsf{T} \in \mathbb{C}^4.$$

Let $\{\beta_1, \beta_2\}$ denote an integral basis over $\mathbb{Z}[i]$ for some ideal $\mathcal{I} \subseteq \mathcal{O}_{\mathbb{L}}$. Then, in accordance with (5), $\mathbf{X}^c$ represents a point in the (complex) lattice whose generator matrix is given by

$$\mathbf{G}^c = \begin{bmatrix} \beta_1 & \beta_2 & 0 & 0 \\ 0 & 0 & \beta_1 & \beta_2 \\ 0 & 0 & \gamma\sigma(\beta_1) & \gamma\sigma(\beta_2) \\ \sigma(\beta_1) & \sigma(\beta_2) & 0 & 0 \end{bmatrix}, \tag{7}$$

i.e.,

$$\mathbf{x}^c = \mathbf{G}^c \left[ a_1 \ a_2 \ a_3 \ a_4 \right]^{\mathsf{T}}, \ \{a_i\}_{i=1}^4 \in \mathbb{Z}(\imath).$$

The corresponding real lattice generator matrix is given by

$$\mathbf{G} = \begin{bmatrix} \mathrm{Re}(\mathbf{G}^c) & -\mathrm{Im}(\mathbf{G}^c) \\ \mathrm{Im}(\mathbf{G}^c) & \mathrm{Re}(\mathbf{G}^c) \end{bmatrix}.$$

It is now evident that the choice of parameters $\gamma$ and $\{\beta_1, \beta_2\}$ completely determines the lattice structure of the ST code (assuming a particular generator $\sigma$ for the group of automorphisms). Furthermore, the choice of these parameters in conjunction with (5) amounts to the choice of a particular subset $\mathcal{L}$ of $\mathcal{O}_{\mathbb{L}}$ to be the signaling alphabet. The key to ensuring good constellation shaping lies in an intelligent choice of the non-norm element and the integral basis.

In [4], these parameters are chosen to ensure that the resultant lattice generated by $\mathbf{G}$ is a rotated version of the cubic lattice $\mathbb{Z}^{2MT}$, i.e., that $\mathbf{G}$ is a unitary matrix. The cubic shaping is in fact the best possible shaping that we can obtain by a linear encoder over the reals (linear-dispersion code). No shaping gain can be achieved by a linear map: at most, the encoder does not increase the transmit energy. This is indeed obtained by $\mathbf{G}$ unitary, that is an isometry of $\mathbb{R}^{2MT}$. The authors in [4] provide such constructions for $2 \times 2$, $3 \times 3$, $4 \times 4$ and $6 \times 6$ (square) ST codes with NVD and have termed the resultant ST codes as *perfect codes*. More recently, [5] presented perfect ST code constructions for arbitrary number of transmit antennas and also for the rectangular case ($T \geq M$).

### B. The S-LaST Construction

We wish to obtain LaST codes with the following properties:

1) the NVD property;

2) the underlying lattice $\Lambda_c$ (referred to as the coding lattice in the following) has large fundamental coding gain $\gamma_c(\Lambda_c)$ (see (6));

3) the shaping region $\mathcal{R}$ is as close as possible to a sphere.

We term the resulting codes as Structured-LaST (S-LaST) codes. The third property yields good *shaping gain* $\gamma_s$, defined as the ratio of the normalized second moment of an $n$-dimensional hypercube to that of the shaping region $\mathcal{R}$. If the shaping region is an $n$-dimensional hypercube, as in the case of perfect codes, then $\gamma_s = 1$. Choosing a better shaping region $\mathcal{R}$ does not change the geometric arrangement of the lattice points, but the average transmitted energy is decreased thanks to shaping. The above three requirement are simultaneously achieved using a nested lattice (Voronoi) construction and a non-linear modulo-lattice encoder nicknamed *sphere encoder*.[3]

Let $\mathbf{G}_p$ denote the generator matrix of a perfect code (unitary), and let $\mathbf{G}_\Lambda$ denote the generator matrix of a good $2MT$-dimensional integral lattice $\Lambda$, that is, a lattice with large fundamental coding gain (such lattices are available in the literature [6]). Define $\Lambda_c$ to be the lattice with generator matrix $\mathbf{G}_{\Lambda_c} = \mathbf{G}_p \mathbf{G}_\Lambda$ and let $\Lambda_s$ (referred to as the shaping lattice) be a sublattice of $\Lambda_c$ such that $\Lambda_s$ has good shaping gain. Let $[\Lambda_c/\Lambda_s]$ denote the nesting ratio, that is, the cardinality of the quotient group $\Lambda_c/\Lambda_s$.

Then, we construct a structured LaST code $\mathcal{X}$ as the set of all distinct points $\mathbf{x}$ given by

$$\mathbf{x} = [\boldsymbol{\lambda} + \mathbf{u}_0] \mod \Lambda_s$$

as $\boldsymbol{\lambda}$ varies in $\Lambda_c$, and $\mathbf{u}_0$ is a translation vector used to symmetrize the code.

Although not necessary, in all cases considered in this paper we let $\Lambda_s = Q\Lambda_c$, $Q \in \mathbb{Z}^+$ for simplicity, i.e., we use a self-similar shaping lattice. The rationale behind this choice is that it is well-known that for moderate dimensions, the best lattices with respect to coding gain are also good quantizers, i.e., have good shaping gain. The coding rate is given by $R = \frac{1}{T} \log[\Lambda_c/\Lambda_s] = 2M \log Q$. Notice also that because of the "rotation" matrix $\mathbf{G}_p$ and the fact that $\Lambda$ is an integral lattice, the set of points $\mathcal{X}$ represented as complex matrices has the NVD property.

*Theorem 1:* The space-time code $\mathcal{X}$ derived from the lattice $\mathbf{G}_{\Lambda_c} = \mathbf{G}_p \mathbf{G}_\Lambda$ using a nested-lattice structure corresponds to a space-time code derived from CDA with non-vanishing determinant and hence achieves the optimal DMT over any fading channel statistics.

*Proof:* Recall that $\mathbf{G}_p$ corresponds to a ST code with NVD, i.e., the set of all non-zero lattice vectors $\mathbf{z} \in \mathbf{G}_p\mathbb{Z}^{2MT}$, represented as complex matrices $\mathbf{Z}^c$, have $\det[\mathbf{Z}^c(\mathbf{Z}^c)^{\mathsf{H}}]$ bounded away from zero by some constant term $\mathrm{SNR}^0$ (up to order of exponent of SNR). Since $\Lambda$ is an integral lattice, there exists a $k \in \mathbb{R}$ such that $k\mathbf{G}_\Lambda$ generates a sublattice of $\mathbb{Z}^{2MT}$. It follows that the LaST code $k\mathcal{X}$ generated by $k\mathbf{G}_p\mathbf{G}_\Lambda$ is a sublattice of $\mathbf{G}_p\mathbb{Z}^{2MT}$ and therefore satisfies

$$\min_{\mathbf{X} \in \mathcal{X}: \ \mathbf{X} \neq 0} \det(\mathbf{X}\mathbf{X}^{\mathsf{H}}) \doteq k^{-2M}\mathrm{SNR}^0 \doteq \mathrm{SNR}^0.$$

---

[3]Tree-search algorithms to perform the Closest Lattice Point Search (CLPS), based on Pohst enumeration [26] and generalized in [22], [23] are generally nicknamed "sphere decoders" if used for minimum distance lattice decoding or "sphere encoders" if used for modulo-lattice precoding, in the current communication and coding theoretic literature. The reason of the nickname follows from the bounded-distance enumerative decoding of the Pohst lattice point enumeration and variants thereof.

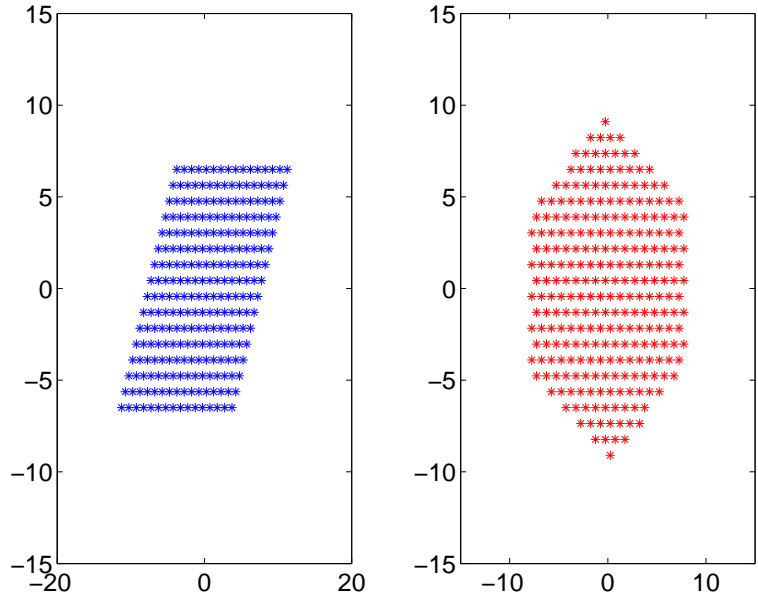

Fig. 1.   Illustrating the Sphere-Encoder: Hexagonal Lattice, $Q = 16$, linear map (left) and sphere-encoded map (right)

The proof of DMT optimality now follows from [2], [3].                                                      ∎

The modulo-$\Lambda_s$ "sphere-encoder" is easily implemented by some CLPS, using some "sphere decoding" algorithm [22], [23]. The shaping effect of *sphere-encoding* is best illustrated using a 2-dimensional example. Suppose that $\Lambda_c$ is the hexagonal lattice in two dimensions. Set $Q = 16$. The constellations corresponding to the linear map (centred at the origin) and the sphere-encoder are shown in Fig. 1. As the value of $Q$ increases, the sphere-encoded constellation fills the fundamental Voronoi region of the hexagonal lattice uniformly. Although both constellations correspond to signalling from the hexagonal lattice, the energy saving of the sphere-encoder is evident.

Example 1: (**The Golden-Gosset S-LaST code**) When $M = 2$, we choose $\mathbf{G}_p$ to be the lattice generator matrix of the Golden code [11] and $\mathbf{G}_\Lambda$ to be the generator matrix of the Gosset lattice $E_8$, which are respectively given by

$$\mathbf{G}_p = \frac{1}{\sqrt{5}} \begin{bmatrix} \mathrm{Re}(G_p^c) & -\mathrm{Im}(G_p^c) \\ \mathrm{Im}(G_p^c) & \mathrm{Re}(G_p^c) \end{bmatrix},$$

where

$$G_p^c = \begin{bmatrix} \eta & \theta\eta & 0 & 0 \\ 0 & 0 & \eta & \theta\eta \\ 0 & 0 & \gamma\sigma(\eta) & \gamma\sigma(\theta)\sigma(\eta) \\ \gamma\sigma(\eta) & \gamma\sigma(\theta)\sigma(\eta) & 0 & 0 \end{bmatrix},$$

$\theta = \frac{1+\sqrt{5}}{2}$, $\sigma(\theta) = 1 - \theta$, $\eta = 1 + \imath - \imath\theta$, $\sigma(\eta) = 1 + \imath - \imath\sigma(\theta)$, $\gamma = \imath$, and

$$\mathbf{G}_\Lambda = \begin{bmatrix} 2 & -1 & 0 & 0 & 0 & 0 & 0 & 0.5 \\ 0 & 1 & -1 & 0 & 0 & 0 & 0 & 0.5 \\ 0 & 0 & 1 & -1 & 0 & 0 & 0 & 0.5 \\ 0 & 0 & 0 & 1 & -1 & 0 & 0 & 0.5 \\ 0 & 0 & 0 & 0 & 1 & -1 & 0 & 0.5 \\ 0 & 0 & 0 & 0 & 0 & 1 & -1 & 0.5 \\ 0 & 0 & 0 & 0 & 0 & 0 & 1 & 0.5 \\ 0 & 0 & 0 & 0 & 0 & 0 & 0 & 0.5 \end{bmatrix}.$$

*Example 2:* (**The Golden+ Algebra** ($\mathcal{GA}+$) **S-LaST code**) Our second example is based on a $2 \times 2$ ST code derived from a maximal order of a CDA [12]. The Golden+ algebra [12] is defined to be $\mathcal{GA}+ = (\mathbb{Q}(\delta)/\mathbb{Q}(\imath), \sigma, \imath)$, where $\delta$ is the first quadrant square root of $2 + \imath$ and the automorphism $\sigma$ is determined by $\sigma(\delta) = -\delta$. The maximal order $O$ of $\mathcal{GA}+$ is generated by the following ordered $\mathbb{Z}(\imath)$−basis:

$$\left\{ \begin{bmatrix} 1 & 0 \\ 0 & 1 \end{bmatrix}, \begin{bmatrix} 0 & 1 \\ \imath & 0 \end{bmatrix}, \frac{1}{2}\begin{bmatrix} \imath + \imath\delta & \imath - \delta \\ -1 + \imath\delta & \imath - \imath\delta \end{bmatrix}, \frac{1}{2}\begin{bmatrix} -1 - \imath\delta & \imath + \imath\delta \\ -1 + \delta & -1 + \imath\delta \end{bmatrix} \right\}. \tag{8}$$

The Golden+ code [12] corresponds to the left ideal of the maximal order generated by

$$\mathbf{M} = \begin{bmatrix} (1-\delta)^3 & 0 \\ 0 & (1+\delta)^3 \end{bmatrix}. \tag{9}$$

In this case, we choose $\mathbf{G}_\Lambda$ to be the lattice generator matrix corresponding to this left ideal of the maximal order and $\mathbf{G}_p = \mathbf{I}$ (trivial rotation). Notice that this choice does not maximize the fundamental coding gain (the Golden-Gosset S-LaST code has a higher density), but the minimum determinant of the Golden+ S-LaST code is better than that of the Golden-Gosset code. It is a priori not clear which effect will dominate the performance in terms of error probability; this will be answered in the simulation results to follow.

*C. Performance under low-complexity MMSE-GDFE Lattice Decoding*

Unfortunately, due to the usage of a non-linear encoding to achieve shaping gain, ML decoding of the resulting code is very complicated, requiring essentially the exhaustive enumeration of the whole codebook. Notice that a similar problem arises in the case of the $\mathcal{GA}+$ code in [12], where linear encoding would result in very bad shaping. The authors in [12] have obtained shaping by enumerating the minimum energy codewords and perform exhaustive decoding, both these are feasible only for low spectral efficiencies.

Hence, we resort to suboptimal MMSE-GDFE lattice decoding (see [1], [22] for details). It has been proven that this decoder achieves the optimal DMT in the random coding sense, for a specific ensemble of random lattices. Here, we use it with our deterministic non-random constructions. We do not claim that the resulting schemes achieve the optimal DMT under lattice decoding. Nevertheless, the performance of these codes is outstanding. In our simulations, we make use of a random translation vector $\mathbf{u}_0$, uniformly distributed over a very large hypercube with volume much larger than the volume of the shaping region. This random "dithering" is known to the receiver, and is subtracted before decoding, as explained in [1]. With this "trick", we ensure that the transmitted points have energy exactly equal to the second moment of $\Lambda_s$ and have exactly zero mean. Furthermore, dithering symmetrizes the scheme and makes the error probability independent of the transmitted codeword.

Fig. 2 compares the performance of two $2 \times 2$ ST codes derived from CDA with $R = 16$ bpcu and $N = 2$. The two ST codes chosen in this case have $\gamma_c(\Lambda_c)$ equal to $0.8365$ and $1.4142$ respectively. Sphere encoding and MMSE-GDFE lattice decoding are used in both cases. We notice about one dB of gain due to better fundamental coding gain of the lattice.

In order to illustrate the benefit of constellation shaping, we plot in Fig. 3 the performance of a $(2 \times 2)$ ST code derived from CDA first using linear encoding of the information symbols and ML decoding and then using sphere encoding and MMSE-GDFE decoding ($R = 16$ bpcu, $N = 2$). The particular ST code chosen has $\gamma_c(\Lambda_c) = 0.8365$. Quite a significant gain of about 3.5 dB results from codebook shaping in this particular case.

For the case of $M = 2$, we compare the performance of the Golden Code [11], which is a perfect $2 \times 2$ ST code (with $\gamma_c(\Lambda_c) = 1$), with the Golden-Gosset $2 \times 2$ S-LaST code from Example 1, ($\gamma_c(E_8) = 2$). Fig. 4 shows plots of the Golden code under ML decoding and MMSE-GDFE lattice decoding in comparison with the Golden-Gosset S-LaST code with MMSE-GDFE lattice decoding at rates of $4$ and 16 bpcu. At 4 bpcu, the (real) information symbol constellation corresponds to BPSK signaling on each dimension ($Q = 2$). In this case, the signal points of the Golden code in 8-dimensional space lie on the surface of a sphere (they are vertices of the rotated hypercube). Therefore, the $2 \times 2$ perfect code construction is optimal for 4 bpcu also in terms of shaping. This intuition is verified by the plots corresponding to 4 bpcu in Fig. 4. However, when the number of bits per channel use increases, the effect of the coding gain of the lattice and the shaping gain begin to show up. At 16 bpcu, the Golden-Gosset S-LaST code with MMSE-GDFE lattice decoding (marginally) outperforms the Golden code with ML decoding (see Fig. 4). These plots also serve to illustrate that MMSE-GDFE lattice decoding is near-ML in performance, while offering significant reductions in complexity.

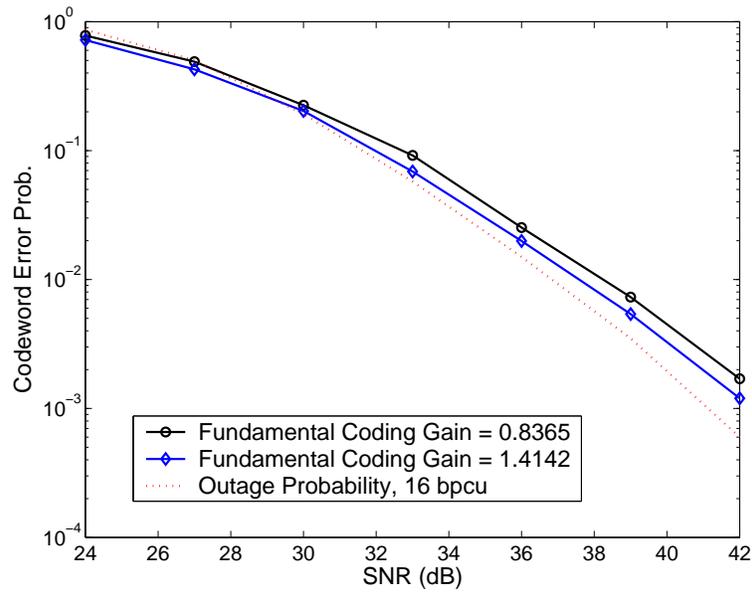

Fig. 2.   Effect of fundamental coding gain on performance: $2 \times 2$ ST codes derived from CDA, 16 bpcu, $N = 2$, MMSE-GDFE lattice decoding

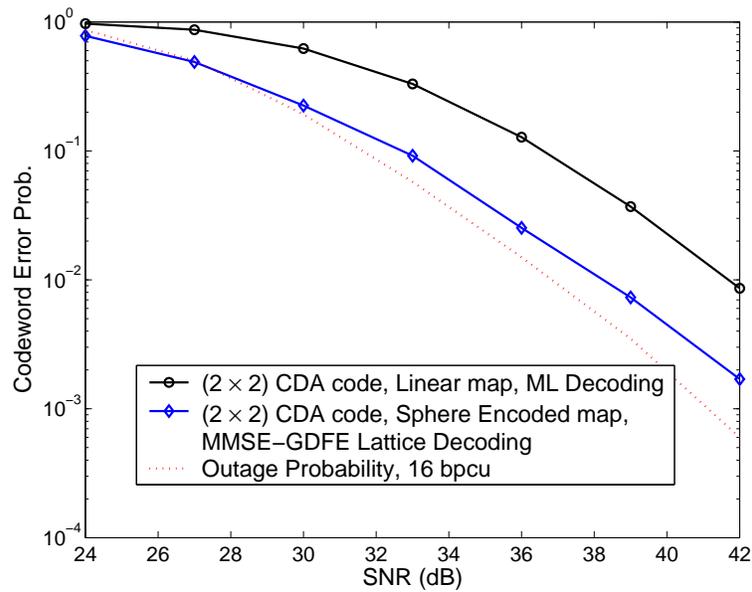

Fig. 3.   Effect of shaping gain on performance: $2 \times 2$ ST code derived from CDA, 16 bpcu, $N = 2$

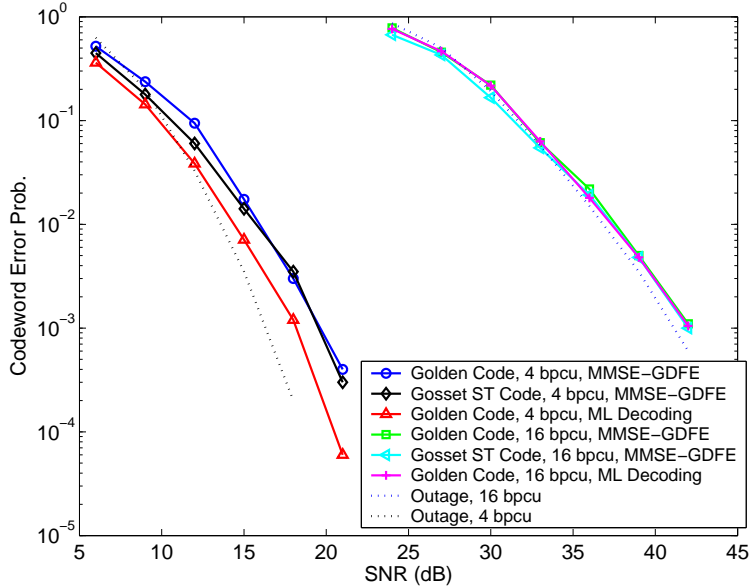

Fig. 4.   Comparing the Golden Code with the Rotated Gosset Lattice ST Code, $N = 2$

In Fig. 5, we present comparisons of the Golden code with ML decoding, the Golden-Gosset S-LaST code (see Example 1) and the $\mathcal{GA}+$ S-LaST code (see Example 2), at 16 bpcu. While the fundamental coding gain of the lattice corresponding to the $\mathcal{GA}+$ code is less than the coding gain of $E_8$, the loss in density is compensated for by an increase in the minimum determinant. Both the Golden-Gosset and the $\mathcal{GA}+$ S-LaST codes with MMSE-GDFE lattice decoding outperform the Golden code with ML decoding.

For the $3 \times 3$ case, we compare the performance of two perfect codes from [5] and [4] (with base alphabets QAM and HEX respectively) with an S-LaST code based on a rotated version of the $\Lambda_{18}$ lattice, which is the best known lattice packing in 18-dimensions [6]. MMSE-GDFE lattice decoding is used for all cases. The results shown in Fig. 6 show a significant gain for both 6 and 24 bpcu resulting from the increased lattice coding gain and shaping.

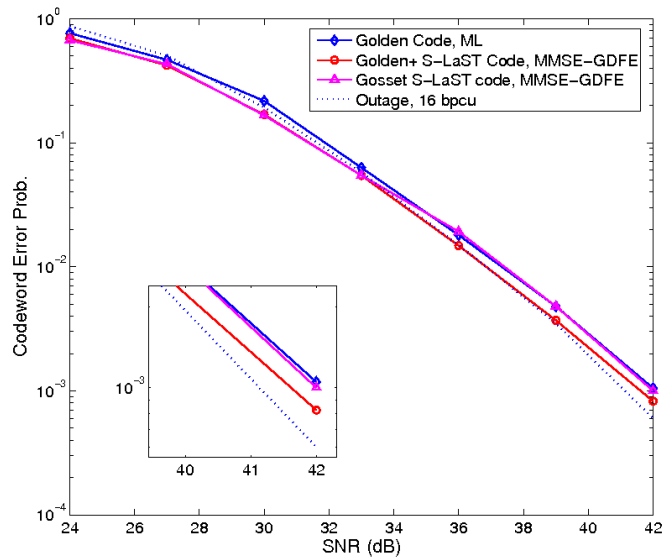

Fig. 5.  Performance of the $2 \times 2$ Golden code, Golden-Gosset and $\mathcal{GA}+$ S-LaST codes at $R = 16$ bpcu. The inset shows a portion of the plot zoomed for clarity.

In Fig. 7 we compare the performance of the $2\times2$ Golden-Gosset S-LaST code ($T = 2$) with rectangular $2 \times 4$ and $2 \times 6$ S-LaST codes constructed using the horizontal-stacking construction [2] in conjunction with the Barnes-Wall ($\Lambda_{16}$) ($\gamma_c(\Lambda_{16}) = 2.8284$) and Leech ($\Lambda_{24}$) ($\gamma_c(\Lambda_{24}) = 4$) lattices respectively. The length-24 cyclic code $\mathcal{G}_{24}(Z_4)$ constructed in [10] was used to construct an isomorphic version of the Leech lattice using construction-A [6]. MMSE-GDFE lattice decoding is used for all three ST codes. In accordance with intuition, the performance approaches outage probability as $T$ increases, owing to better values of $\gamma_c(\Lambda_c)$.

## IV. THE S-LaST TCM SCHEME

Motivated by the fact that in practical wireless communications $M$ is limited by transmitter complexity to be a small integer (typically 2 or 4, in current IEEE802.11n MIMO extension of wireless local area networks) while $T$ may be of the order of 100 channel uses, our objective in this section is to construct $M \times T$ ST codes for the case of $T \gg M$. For ease of exposition and without loss of fundamental generality, we will focus on the case where $T = LM$, for some integer $L$. TCM has the nice feature that a single trellis code can generate any desired block length, with decoding complexity linear in $L$, using a Viterbi decoder. Furthermore, the construction of TCM schemes is rather well understood and a rich literature exists for the Gaussian channel (see [13], [14], [15] and references therein), the scalar fading channel (see [16] and references therein) and for the MIMO fading channel [17], [18], [19].

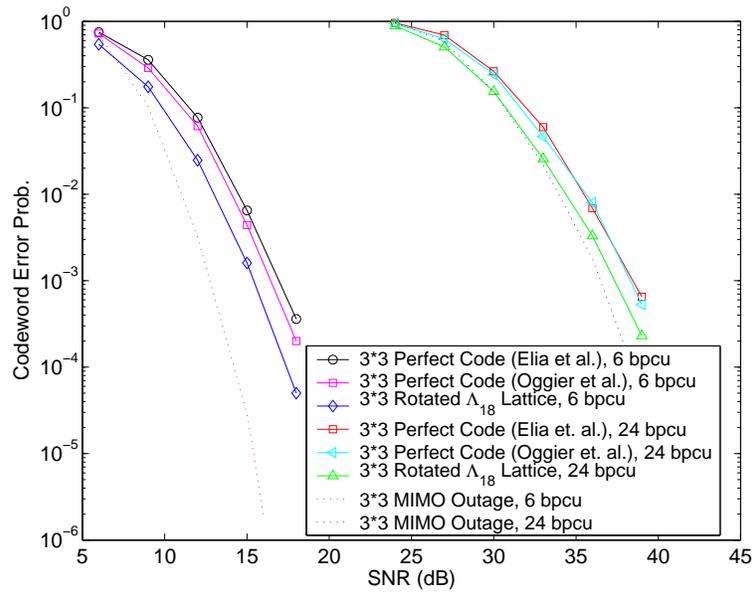

Fig. 6.   $3 \times 3$ ST Codes under MMSE-GDFE lattice decoding, $N = 3$

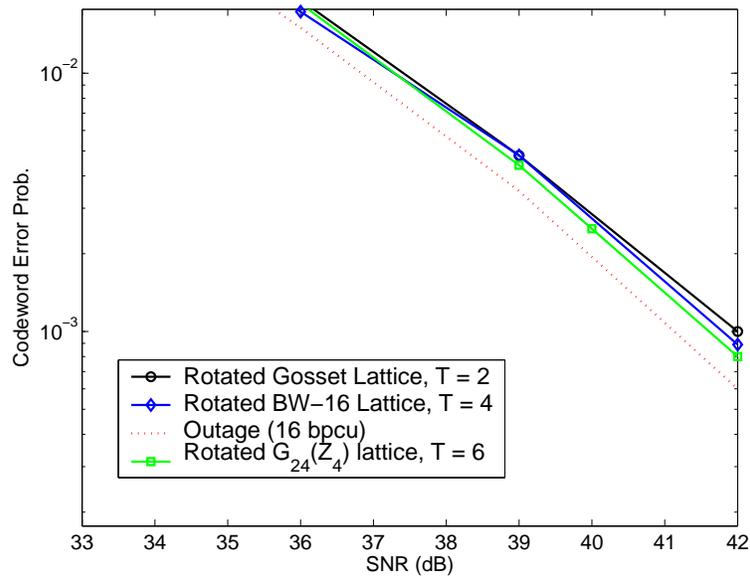

Fig. 7.   Increasing the Coding Length, $M = N = 2$, $T = 2, 4, 6$, $R = 16$ bpcu, MMSE-GDFE lattice Decoding

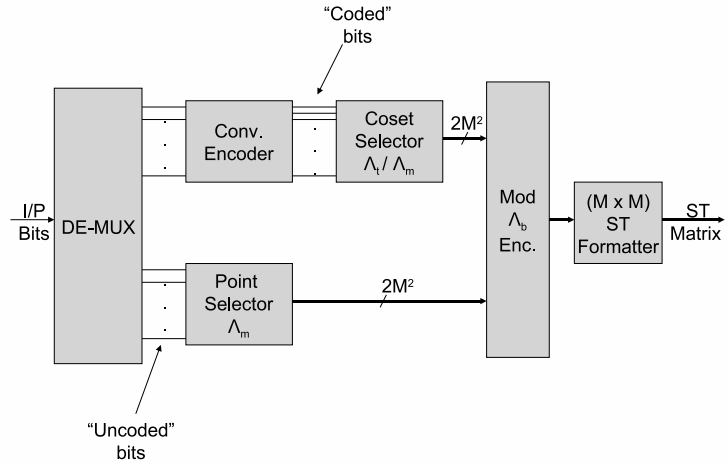

Fig. 8.   S-LaST TCM Encoder

### A. Encoder

Consider a three level partition $\Lambda_t \supset \Lambda_m \supset \Lambda_b$ (where the subscripts indicate 'top', 'middle' and 'bottom') of lattices in $\mathbb{R}^n$, with $n = 2M^2$. Let $[\Lambda_t/\Lambda_m] = \mathcal{M}$ and let the cosets of $\Lambda_m$ in $\Lambda_t$ be indicated by $\mathcal{C}_i \triangleq \{\mathbf{v}_i + \Lambda_m\}$, for $i = 1, \dots, \mathcal{M}$, where each $\mathbf{v}_i$ is a coset representative of $\mathcal{C}_i$. From each coset $\mathcal{C}_i$, we carve a finite set of $\mathcal{N}$ points, denoted by $\{\mathbf{v}_i + \mathbf{c}_j : \mathbf{c}_j \in \Lambda_m, j = 1, \dots, \mathcal{N}\}$. These points are chosen via a modulo-$\Lambda_b$ sphere encoder, that will be described in the following. Also, we choose $\Lambda_b$ such that $\mathcal{N} = [\Lambda_m|\Lambda_b]$. In all the examples presented here, we use $\Lambda_b = Q\Lambda_m$, for some $Q \in \mathbb{Z}^+$ (i.e., we use again a self-similar shaping lattice). In this case, $\mathcal{N} = Q^{2M^2}$.

We make use of Forney's general "coset coding" framework [8]. A block diagram of the encoder is shown in Fig. 8. During each block $k = 1, \dots, L$ comprising of $M$ channel uses each, a block of $(\log \mathcal{M})/r + \log \mathcal{N}$ information bits enters the encoder. The top $(\log \mathcal{M})/r$ information bits are input to a convolutional encoder of (binary) rate $r$, that outputs $\log \mathcal{M}$ coded bits, which select the index $i_k \in \{1, \dots, \mathcal{M}\}$ of a coset in $\Lambda_t/\Lambda_m$. The remaining $\log \mathcal{N}$ information bits select the index $j_k$ of a point in the finite constellation carved from the selected coset $\mathcal{C}_{i_k}$.

The transmitted vector at time $k$ is given by

$$\mathbf{x}_k \quad = \quad [\mathbf{c}_{j_k} + \mathbf{v}_{i_k} + \mathbf{u}_k] \mod \Lambda_b \tag{10}$$

where $\mathbf{u}_k$ is an optional random dithering signal known to the receiver, that serves to symmetrize the overall TCM code and to induce the uniform error property. The vector $\mathbf{x}_k$ is then mapped into an $M \times M$ complex matrix and transmitted in $M$ channel uses across the MIMO channel. The rate of the S-LaST TCM scheme is given by

$$R = \frac{(\log \mathcal{M})/r + \log \mathcal{N}}{M} \text{ bits/channel use.}$$

It should be noticed that $\mathbf{x}_k = \mathbf{c}_{j_k} + \mathbf{v}_{i_k} + \mathbf{u}_k - \boldsymbol{\lambda}_k$ for some $\boldsymbol{\lambda}_k \in \Lambda_b$ that is a function of $\mathbf{c}_{j_k}, \mathbf{v}_{i_k}, \mathbf{u}_k$. Further, $\mathbf{x}_k \in \mathcal{V}(\Lambda_b)$. Since $[\Lambda_m | \Lambda_b] = \mathcal{N}$, the mapping between the uncoded bits and the constellation points in each coset is one-to-one.

## B. Decoder

The (real equivalent) received point at each block $k$ is given by

$$\mathbf{y}_k = \mathbf{H}\mathbf{x}_k + \mathbf{w}_k,$$

for $k = 1, \ldots, L$. In general, the trellis of the S-LaST TCM scheme has $\mathcal{N}$ parallel transitions per trellis branch, corresponding to the $\mathcal{N}$ points in the intersection $\mathcal{C}_i \cap \mathcal{V}(\Lambda_b)$, on each branch labeled by the coset $\mathcal{C}_i$. Consider time $k$, and a branch labeled by coset $\mathcal{C}_i$. The corresponding branch metric for a ML trellis decoder (implemented via the Viterbi algorithm) is given by

$$B_{i,k} = \min_{\mathbf{c} \in \Lambda_m \cap \mathcal{V}(\Lambda_b)} |\mathbf{y}_k - \mathbf{H}(\mathbf{v}_i + \mathbf{c} + \mathbf{u}_k)|^2. \tag{11}$$

Computing this branch metric amounts to exhaustive enumeration of all points of $\Lambda_m$ in the Voronoi region $\mathcal{V}(\Lambda_b)$ of the shaping lattice.

Since exhaustive enumeration is usually too complex, we resort once again to a suboptimal MMSE-GDFE lattice decoder along the lines of [1], in order to compute an approximate ML branch metric for the Viterbi decoder. First, we relax the minimization in (11) to take into account all points of $\Lambda_m$ (Lattice decoding), i.e., we consider the suboptimal branch metric

$$B_{i,k} = \min_{\mathbf{c} \in \Lambda_m} |\mathbf{y}_k - \mathbf{H}(\mathbf{v}_i + \mathbf{c} + \mathbf{u}_k)|^2. \tag{12}$$

This amount to solving a CLPS problem for the channel-modified lattice $\mathbf{H}\Lambda_m$, with respect to the point $\mathbf{y}_k - \mathbf{H}(\mathbf{v}_i + \mathbf{u}_k)$, where $\mathbf{u}_k$ is a known dithering vector and $\mathbf{v}_i$ depends on the label of the branch for which we compute the metric. The surviving path among the parallel paths corresponds to the argument $\mathbf{c}$ that minimizes (12).

Then, we further modify the suboptimal metric following the MMSE-GDFE paradigm (see [1] for the details). Let $\mathbf{F}$ and $\mathbf{B}$ denote the forward and backward filters of the MMSE-GDFE as defined in [1]. At each time $k$, the receiver obtains the following set of modified channel observations

$$\mathbf{y}'_{i,k} = \mathbf{F}\mathbf{y}_k - \mathbf{B}(\mathbf{v}_i + \mathbf{u}_k), \ 1 \le i \le \mathcal{M}.$$

Using the properties of the matrices $\mathbf{F}$ and $\mathbf{B}$, these can be written as

$$\begin{aligned}
\mathbf{y}'_{i,k} &= \mathbf{F}\left[\mathbf{H}(\mathbf{c}_{j_k} + \mathbf{v}_{i_k} + \mathbf{u}_k - \boldsymbol{\lambda}_k) + \mathbf{w}_k\right] - \mathbf{B}[\mathbf{u}_k + \mathbf{v}_i] \\
&= \mathbf{B}(\mathbf{c}_{j_k} + \mathbf{v}_{j_k} - \boldsymbol{\lambda}_k - \mathbf{v}_i) - [\mathbf{B} - \mathbf{FH}](\mathbf{c}_{j_k} + \mathbf{v}_{i_k} - \boldsymbol{\lambda}_k + \mathbf{u}_k) + \mathbf{F}\mathbf{w}_k \\
&= \mathbf{B}(\mathbf{c}_{j_k} + \mathbf{v}_{j_k} - \boldsymbol{\lambda}_k - \mathbf{v}_i) - [\mathbf{B} - \mathbf{FH}]\mathbf{x}_k + \mathbf{F}\mathbf{w}_k \\
&\triangleq \mathbf{B}(\mathbf{c}_{j_k} + \mathbf{v}_{j_k} - \boldsymbol{\lambda}_k - \mathbf{v}_\ell) + \mathbf{e}'_k.
\end{aligned}$$

Notice that $\mathbf{x}_k$ is uniformly distributed over $\mathcal{V}(\Lambda_b)$ and is hence independent of $\mathbf{c}_{j_k}$ and $\mathbf{v}_{j_k}$ [1]. It can be shown that the noise plus self-noise vector $\mathbf{e}'_k$ has the same covariance matrix of the original noise $\mathbf{w}_k$, although it is generally non-Gaussian. Also, $\mathbf{v}_{i_k} - \mathbf{v}_i = 0$ (i.e., it belongs to $\Lambda_m$) if $i_k = i$, while it belongs to some coset of $\Lambda_m$ in $\Lambda_t$ not equal to $\Lambda_m$ if $i_k \ne i$.

For each branch labeled by coset $\mathcal{C}_i$, the low-complexity Viterbi decoder computes branch metric

$$B_{i,k} = \min_{\mathbf{z} \in \mathbb{Z}^{2M^2}} \left| \mathbf{y}'_{i,k} - \mathbf{B}\mathbf{G}_{\Lambda_m}\mathbf{z} \right|^2$$

where $\mathbf{G}_{\Lambda_m}$ denotes a generator matrix for $\Lambda_m$. This can be obtained by a sphere decoder applied to the channel-modified lattice $\mathbf{B}\Lambda_m$. It is clear that the branch metric for the correct coset (i.e., for $i = i_k$) will be smaller than the branch metric for an incorrect coset, with high probability.

## C. Construction of suitable lattice partition chains

In order to ensure good performance, we choose the component $M \times M$ code of the S-LaST TCM scheme to be approximately universal. We will therefore choose $\Lambda_t$ to be the lattice corresponding to an ST code derived from CDA with NVD. In order to construct $\Lambda_m$ and $\Lambda_b$, we will first discuss the important special case when $\Lambda_t$ corresponds to a perfect code, and then treat the more general case.

*1) Partitions of perfect codes:* Let $\Lambda_t$ be the lattice corresponding to a perfect code [4], [5], with generator matrix $\mathbf{G}_p$. Then, $\Lambda_t$ is a rotated version of the cubic lattice $\mathbb{Z}^{2M^2}$. Following what was done before for the case of short block codes, we choose $\Lambda_m$ to be the best known integral lattice packing in $2M^2-$dimensional space, rotated by $\mathbf{G}_p$. Also, we set $\Lambda_b = Q\Lambda_m$. For example, when $M = 2$, we choose $\Lambda_m$ to be the Golden Gosset lattice. The resulting code shall be named the Golden-Gosset S-LaST TCM scheme.

*2) S-LaST TCM from maximal orders in CDAs:* We choose $\Lambda_t$ to be the lattice corresponding to the maximal order of a given CDA. An example for the case when $M = 2$ would be the lattice corresponding to the $\mathcal{GA}+$ code that we made use of for the short block-length case in Example 2. Similar to the approach used in [20], [21] for the cubic lattice case, we will use ideals $\beta O$ of the maximal order for the sublattice $\Lambda_m$. The element $\beta$ yielding a good sublattice is obtained through a computer search, that makes use of the following lemma.

*Lemma 2:* Let $D(\mathbb{L}/\mathbb{Q}(i), \sigma, \gamma)$ be a cyclic division algebra of index $n$, and let $O$ denote an order of $D$. If $\beta$ is an element of the order, then

$$[O|\beta O] = |N_r(\beta)^n|^2.$$

*Proof:* Although this lemma is well known to the mathematics community, we provide a sketch of the proof for completeness. Consider any $\beta \in O$. Then $\beta$ induces a transformation on $O$ with image $\beta O$. These are finitely generated free modules over $\mathbb{Z}$, and so the index of partition is just the determinant of $\beta$ in this action.

We may compute the determinant over the corresponding field. $D$ has rank $2n^2$ over $\mathbb{Q}$. First viewing $D$ as a (right) vector space of dimension $n^2$ over $\mathbb{Q}(i)$, we see that the determinant of multiplication by $\beta$ is $N_r(\beta)^n$. We then apply the norm from $\mathbb{Q}(i)$ to $\mathbb{Q}$ to obtain the determinant. ∎

The computer search performs the following:

1) Fix a desired index of partition $\mathcal{M} = [\Lambda_t|\Lambda_m]$, and a sufficiently large integer $\nu$.

2) Let $O_\nu$ denote the integral closure of $\{-\nu, -\nu + 1, \ldots, \nu - 1, \nu\} \subset \mathbb{Z}$ in $O$. More specifically, if $\gamma_1, \gamma_2, \ldots, \gamma_{2M^2}$ constitutes a basis for $O$ over $\mathbb{Z}$, then

$$O_\nu \triangleq \left\{ \sum_{i=1}^{2M^2} g_i \gamma_i \ \middle| \ -\nu \le g_i \le \nu, \ g_i \in \mathbb{Z} \ \forall \ i \right\}.$$

Notice that such a basis always exists, since every algebraic number field has at least one integral basis [25].

3) For each $\beta \in O_\nu$ that generates a partition with required index $\mathcal{M}$, i.e., satisfying $\left|N_r(\beta)^M\right|^2 = \mathcal{M}$,

compute the fundamental coding gain of the lattice corresponding to $\beta O$, and let $\beta_{\max}$ denote a maximizer.

4) Set $\Lambda_m$ to be the lattice corresponding to $\beta_{\max} O$.

Finally, as before, we use the self-similar shaping lattice $\Lambda_b = Q\Lambda_m$, for some $Q \in \mathbb{Z}^+$.

### D. Code construction examples

In this section, we present two construction examples of S-LaST TCM, the performances of which are compared by simulation.

- The Golden-Gosset S-LaST TCM construction (see Example 1): here $\Lambda_t = \mathbf{G}_p \mathbb{Z}^8$, $\Lambda_m = \mathbf{G}_p E_8$ and $\Lambda_b = Q\Lambda_m$, $Q \in \mathbb{Z}^+$.

- The $\mathcal{GA}+$ S-LaST TCM construction: we choose $\Lambda_t$ to be the lattice corresponding to the $\mathcal{GA}+$ S-LaST code in Example 2. $\Lambda_m$ is obtained using the computer search given above, and corresponds to the left ideal of $\beta^2 O$ generated by $\mathbf{M}$ (given in (9)), where $O$ is the maximal order of the $\mathcal{GA}+$ algebra (see Example 2) and the coordinates of $\beta$ in terms of the ordered basis in (8) are $(-1, -1, 1 - \imath, -1 - \imath)$. We then set $\Lambda_b = Q\Lambda_m$, $Q \in \mathbb{Z}^+$.

Both these codes correspond to a $16-$ary partition $\Lambda_t/\Lambda_m$, as shown in Fig. 9. The minimum determinant increases as one goes down the partition chain. We use the trellis shown in Fig. 10 that is designed such that the transitions leaving/merging into a state have maximum possible minimum determinant.

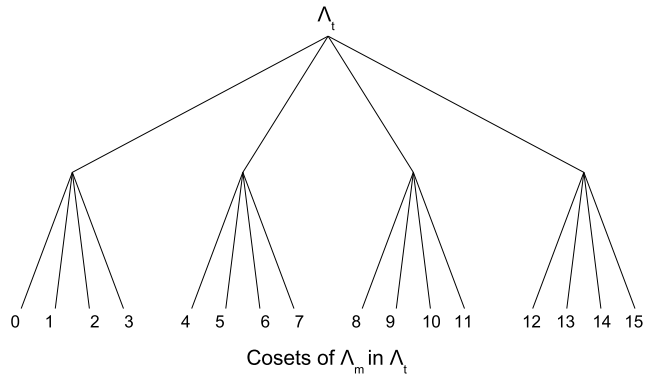

Fig. 9.   Two level partition of the example constructions

In our simulations, we have used block length $T = 260$ channel uses, corresponding to 1300 information bits per packet, at $R = 5$ bpcu. Fig. 11 shows the performance in terms of packet error probability of the above two S-LaST TCM schemes in comparison with the Golden ST TCM (GST-TCM) scheme [21] at 5 bpcu. Also shown is the performance of the "uncoded Golden code" construction [21], which consists of stacking 130 Golden code matrices next to each other (coding is performed only over 2 time-slots). The proposed S-LaST TCM construction is seen to gain around 1 dB over the GST-TCM scheme.

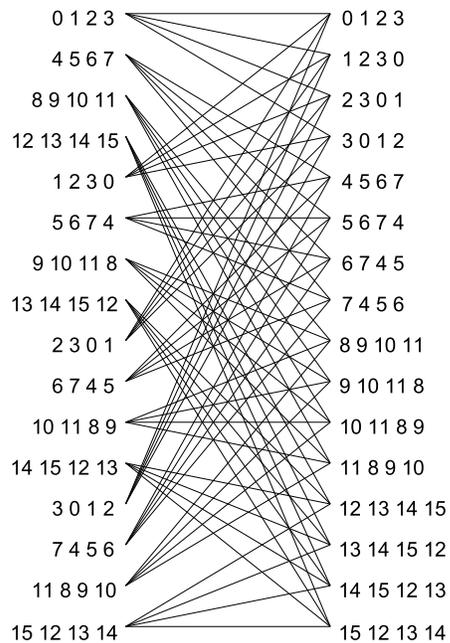

Fig. 10.　16-state trellis used for the example constructions

## V. Conclusions

In this paper, we have advocated the use of structured lattices that are endowed with good packing and shaping properties in the design of space-time codes with both short and long block-lengths. The constructions presented have reasonable decoding complexity, and exhibit excellent performance in terms of error probability.

Quite a few research topics occur naturally as potential follow-up works. While codes with short block-length have performances that are very close to the outage probability, there is still quite a significant gap from outage for the case of long block-lengths. Designing better codes for this scenario remains a challenging open problem. It would also be interesting to explore if there exist better algebraic frameworks that allow us to choose sublattices with good packing and shaping properties.

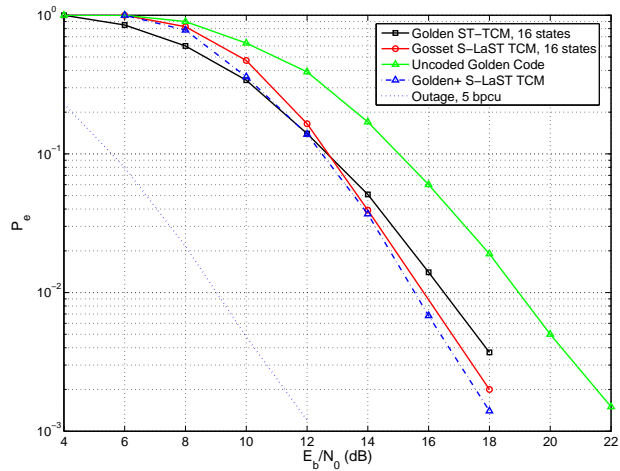

Fig. 11. Performance of the Golden-Gosset and $\mathcal{GA}+$ S-LaST TCM schemes, $R = 5$ bpcu, $T = 260$

## ACKNOWLEDGEMENTS

The authors would like to thank Prof. Robert Guralnick for some useful discussions.